\documentclass[aps,prb,reprint,showpacs]{revtex4-1}
\usepackage{amsmath,amssymb,bm}
\usepackage{latexsym}
\usepackage{subfigure}
\usepackage[pdftex]{graphicx}
\usepackage{color}
\usepackage{hyperref}

\newcommand{\aver}[1]{\langle #1 \rangle}

\newcommand{\CQX}{Cu(Qnx)(Cl$_{(1-x)}$Br$_{x}$)$_{2}$}
\newcommand{\CQC}{Cu(Qnx)Cl$_{2}$}
\newcommand{\CQB}{Cu(Qnx)Br$_{2}$}

\begin{document}

\title{ The tunable quantum spin ladder \CQX}

\author{K. Yu. Povarov}
    \altaffiliation[Previous address: ]{P. L. Kapitza Institute for Physical Problems, RAS, 119334 Moscow, Russia}
    \affiliation{Neutron Scattering and Magnetism, Laboratory for Solid State Physics, ETH Z\"{u}rich, Switzerland}

\author{W. E. A. Lorenz}
    \affiliation{Neutron Scattering and Magnetism, Laboratory for Solid State Physics, ETH Z\"{u}rich, Switzerland}

\author{F. Xiao}
    \altaffiliation[Present address: ]{Department of Physics, Durham University, South Road, Durham, DH1 3LE, United Kingdom}
    \affiliation{Department of Physics, Clark University, Worcester, Massachusetts 01610, USA}

\author{C. P. Landee}
    \affiliation{Department of Physics, Clark University, Worcester, Massachusetts 01610, USA}

\author{Y. Krasnikova}
    \affiliation{P. L. Kapitza Institute for Physical Problems, RAS, 119334 Moscow, Russia}
    \affiliation{Moscow Institute for Physics and Technology, 141700 Dolgoprudny, Russia}

\author{A. Zheludev}
   \affiliation{Neutron Scattering and Magnetism, Laboratory for Solid State Physics, ETH Z\"{u}rich, Switzerland}

\date{\today}

\begin{abstract}
We report magnetic, specific heat and ESR measurements on a series of $S=1/2$ spin ladder compounds \CQX. Down to $T=2$~K all the observables can be described by
the spin ladder model with about 1\% of $S=1/2$ impurities in the background, which are present even in a nominally pure \CQC\ and \CQB, for the whole range of $0\leq x\leq1$. We extract ladder exchange constants $J_{l}(x)$ and $J_{r}(x)$ as well as the gap value $\Delta(x)$ by comparing the experimental data to quantum Monte-Carlo simulations. As the ESR measurements show the $g$-factor of impurities to be precisely equal to $g$-factor of the ladder spins, we suppose the impurities to be the consequence of structural defects resulting in a broken ladder ends.
\end{abstract}



\maketitle

\section{Introduction \label{Introduction}}

The two-leg $S=1/2$ spin ladder is one of the most extensively studied models in the field of low-dimensional magnetism~\cite{GiamarchiBook}. Demonstrating complex properties despite a deceptive simplicity it also benefits from the existence of exact solutions in the limiting cases of non-interacting spin chains and non-interacting dimers. Unfortunately, only a handful of real materials realize the spin-ladder model~\cite{LadderMain, IPA_Haldane, Sulfalane_4leg, DIMPY1}. Among the best examples are organometallic compounds~\cite{Landee_review} such as  the strong-rung material (Hpip)$_{2}$CuBr$_{4}$ with $\alpha\equiv J_{r}/J_{l}\simeq3.9$~\cite{HpipLL, HpipBEC}  and the strong-leg material (C$_{7}$H$_{10}$N)$_{2}$CuBr$_{4}$ also known as DIMPY with $\alpha\simeq0.43$~\cite{DIMPY1, DIMPY_longlive, DIMPY_spectralandtherm}. Potentially the most unusual regime is that of almost equal rung and leg interactions. To date, only two prototype compounds are known to even come close to this "isotropic" ladder model, namely \CQC\ and \CQB\ (CQC and CQB for short)~\cite{ActaCryst1990,Polyhedron2003,Polyhedron2011}. In these materials $\alpha\sim1.7$. An important feature of this material family is the possibility to create generic \CQX\ compounds (CQX) belonging to the same structure type for any $0\leq x\leq1$. As suggested by preliminary studies of these materials with small values of $x$, chemical substitution opens a route to creating spin ladders with continuously tunable exchange constants and magnetic properties~\cite{Polyhedron2011, LandeeInorgChem2012}. In the present work we use a variety of techniques, such as bulk magnetometery, ESR and specific heat measurements, to explore the entire concentration range. We determine the continuous variation of $J_{l}(x)$ and $J_{r}(x)$ for the whole range of Br concentrations $0\leq x\leq1$. In addition, we demonstrate that undesirable effects due to structural disorder and potential randomness of magnetic interactions~\cite{DirtyBoson} are negligible even around $x\sim0.5$ despite the nominally strong disorder. The present finding is unusual, as in most cases through continuous chemical substitution one gets not a continuous tuning of the parameters, but rather two distinct regimes in the vicinity of the pure systems (for example~\cite{CPC_doped}). This makes CQX a unique system where the tuning of the parameters in a continuous manner is possible by halogen substitution.

\section{Experimental details \label{Experiment}}

\subsection{Samples}

\begin{figure}
\begin{center}
  \includegraphics[width=0.4\textwidth]{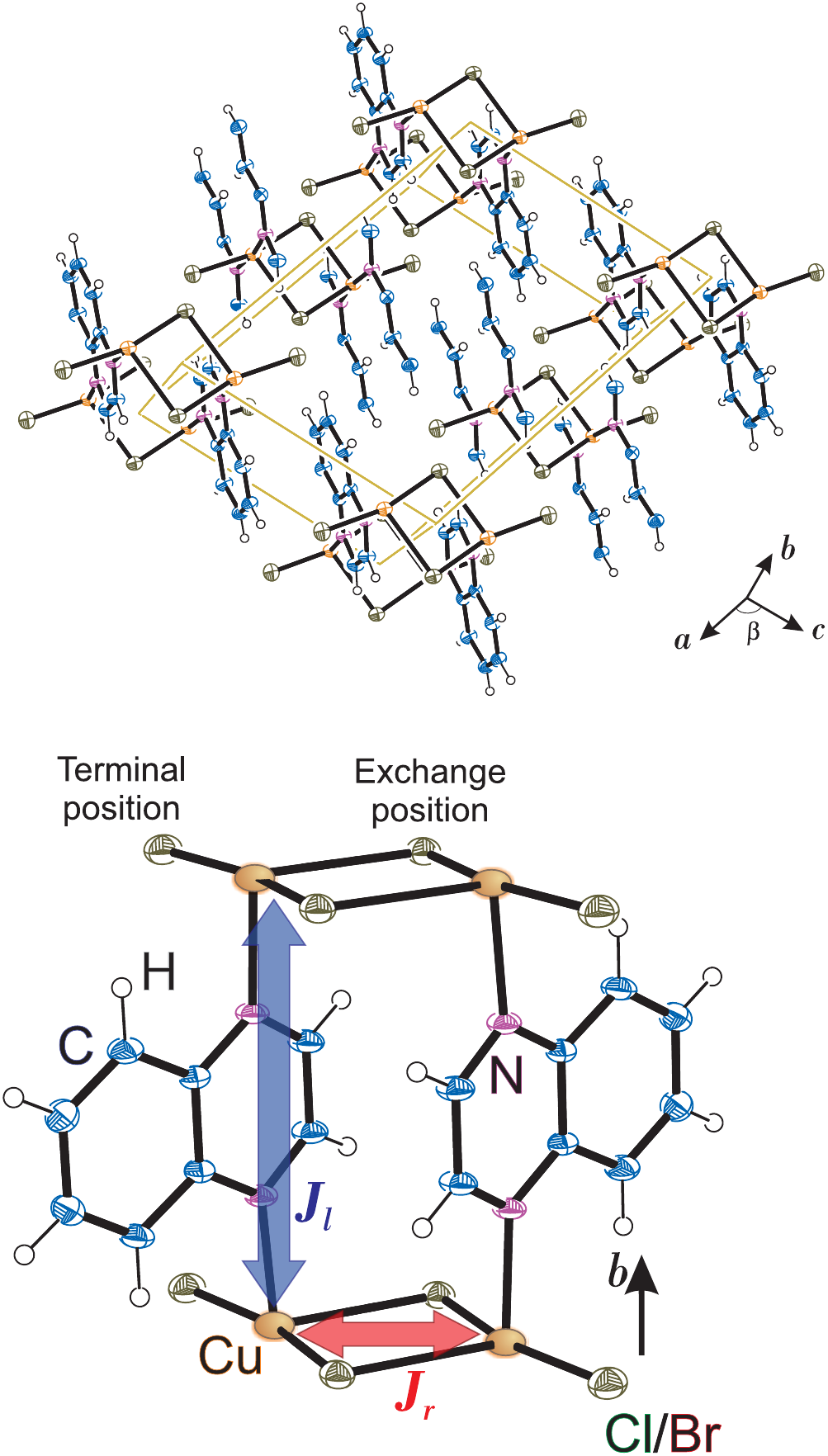}\\
  \caption{Crystal structure of \CQX: monoclinic unit cell with several ladder segments and magnified view on a ladder segment with $J_{l}$ and $J_{r}$ pathways denoted.}\label{FIG:structure}
\end{center}
\end{figure}

The compounds of CQX family belong to the monoclinic $C2/m$ space
group, with lattice parameters $a=13.237$~{\AA}, $b=6.935$~{\AA},
$c=9.775$~{\AA}, $\beta=107.88^{\circ}$ for pure
\CQC~\cite{ActaCryst1990} and $a=13.175$~{\AA}, $b=6.929$~{\AA},
$c=10.356$~{\AA}, $\beta=107.70^{\circ}$ for pure
\CQB~\cite{Polyhedron2003}. Details of the crystal structure can be
found in Figure~\ref{FIG:structure}: Cu$^{2+}$ ions with $S=1/2$ are
bridged together by Qnx molecules, forming chains along the two-fold
rotation axis $b$. Adjacent pairs of chains are in turn coupled via
the bihalide superexchange pathway in the $ac$ plane, resulting in a
ladder configuration. There are four halogen ions per two copper
ions, of which two participate in the rung exchange, and the other
two are in so-called terminal positions. The crystallographic unit
cell includes two equivalent ladder units, related by translation
symmetry.

\begin{figure}
\begin{center}
  \includegraphics[width=0.48\textwidth]{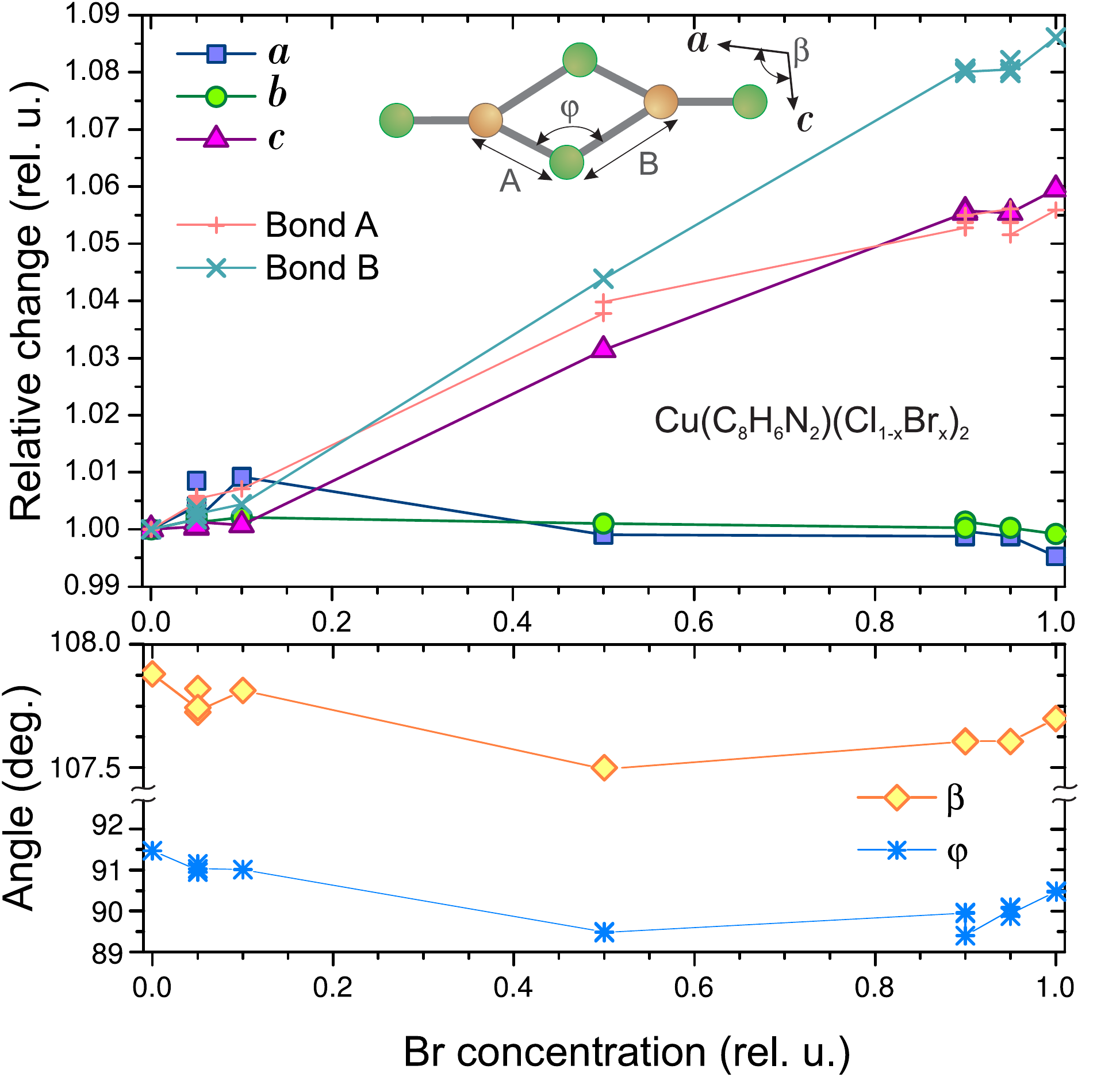}\\
  \caption{Relative variation of lattice ($a$, $b$, $c$, $\beta$) and rung exchange ($A$, $B$, $\varphi$) parameters with Br substitution $x$ as determined by single crystal X-ray diffraction. Parameters for CQC are $a=13.237$~{\AA}, $b=6.935$~{\AA}, $c=9.775$~{\AA}, $\beta=107.88^{\circ}$, $A=2.3$~{\AA}, $B=2.685$~{\AA} and $\varphi=91.47^{\circ}$.}\label{FIG:Cellvariations}
\end{center}
\end{figure}

A series of typically 1~mm$^3$ single crystal CQX samples were synthesized in ETH~Z\"{u}rich~\cite{Crystalgrowth} using slow diffusion in methanol solution, as described in~\cite{Polyhedron2011}.  Nominal substitution $x$ in the series of samples was $x$=0, 0.05, 0.1, 0.15, 0.2, 0.25, 0.4, 0.5, 0.6, 0.75, 0.8, 0.85, 0.9, 0.95, and 1. The crystal structure was verified using single crystal X-ray diffraction for several representative concentrations. As shown in Figure~\ref{FIG:Cellvariations}, one observes a continuous variation of lattice parameters vs. Br content.

For specific heat and ESR measurements single crystals were used as
obtained. For magnetization measurements, where large sample mass is
required, the samples were prepared by orienting powdered CQX
material. For each measurement about 100~mg of CQX crystals were
ground into powder with pestle and mortar and placed together with
40~mg of paraffine into a gelatine capsule. Such a sample was
exposed to $7$~T magnetic field at $T=330$~K for one hour, then
cooled back in field. No further increase in magnetic signal was
observed already after $\sim20$~minutes of exposure to the field. By
comparing the fitted $g$-factors to the maximal values of the
$g$-tensor determined by ESR (see below) we can conclude that the
degree of misalignment in the powdered samples does not overcome
$5^{\circ}$ on average. The direction with maximal $g$ lies
approximately at $45^{\circ}$ to the $a-$axis in the $ac$ plane.

\subsection{Experimental techniques}

Measurements of magnetic properties were made using vibrating sample
magnetometer option for the PPMS system in ETH~Z\"{u}rich. The
oriented powder samples were installed onto a standard PPMS VSM
brass half-tube sample holder with quartz rods and wrapped in PTFE
tape. Measurements of susceptibility and low-temperature
magnetization for each sample were performed during the same
experimental run. ESR measurements were performed in P.~L.~Kapitza
Institute for Physical Problems RAS on a homemade rectangular-cavity
multifrequency spectrometer, designed as an insert to a $^{4}$He
pumping cryostat equipped with superconducting magnet.

\section{Results and discussion \label{Results}}

\subsection{Bulk magnetometry}

\begin{figure}
\begin{center}
  \includegraphics[width=0.5\textwidth]{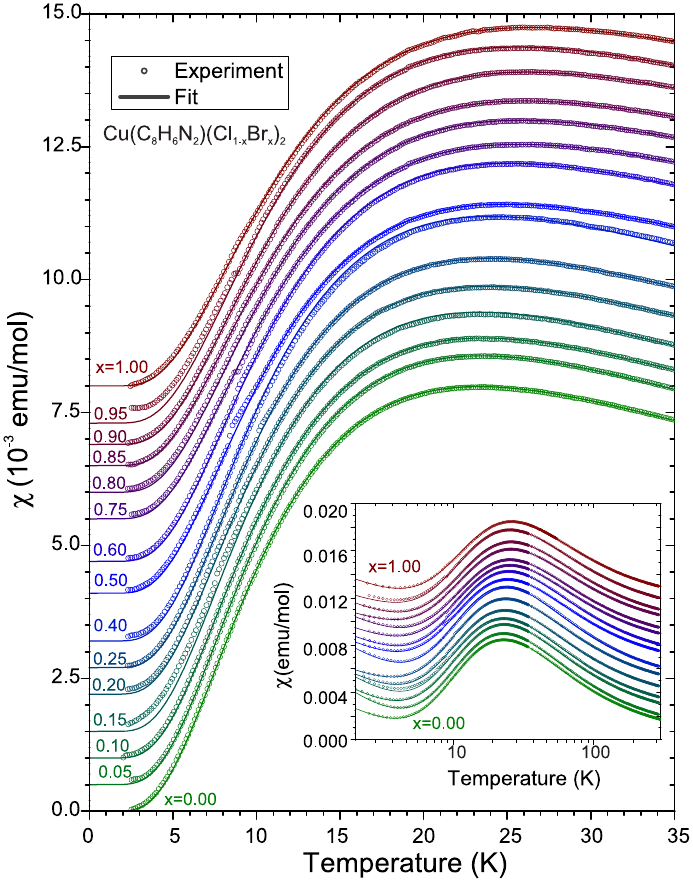}\\
  \caption{Data and fits for magnetic susceptibility of CQX oriented powder, background subtracted. Inset: the same data and fits, including background. In both main figure and inset an offset between different curves is introduced.}\label{FIG:Susceptibility}
\end{center}
\end{figure}

The magnetic susceptibility was obtained as $M(H)/H$ for
$\mu_{0}H=0.1$~T in the field-cooling regime for the whole series of
concentration $x$. For all $x$, the susceptibility curve $\chi(T)$,
depicted in the inset of Figure~\ref{FIG:Susceptibility}, looks
typical for a gapped low-dimensional antiferromagnet. The main
features are Curie--Weiss behaviour at high temperatures and a broad
maximum around 25 -- 30~K. Below this maximum the susceptibility
rapidly decreases. At the lowest temperatures, approaching $2$~K, an
impurity-like Curie tail arises. The rapid decrease of
susceptibility with $T\rightarrow0$ is a signature of a gapped
ground state. No magnetic field or thermal history was observed for
any of the samples.

\begin{figure}
\begin{center}
  \includegraphics[width=0.5\textwidth]{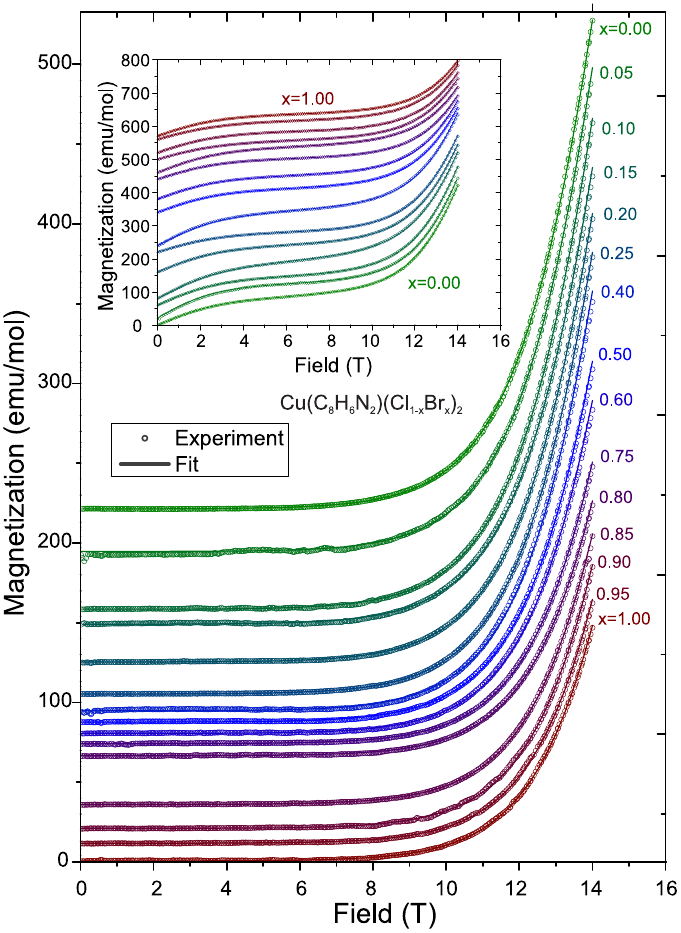}\\
  \caption{Data and fits for the magnetization of CQX oriented powder at $T=2$~K, background subtracted. Inset: the same data and fits including background. In both main figure and inset an offset between different curves is introduced. Colors are the same as in Figure~\ref{FIG:Susceptibility}.}\label{FIG:Magnetization}
\end{center}
\end{figure}

Magnetization measurements were performed at $T=2$~K with field
sweep to $14$~T and back. No hysteresis in the magnetization was
observed. These curves are present in the inset of
Figure~\ref{FIG:Magnetization}. The low-field part of magnetization
curve again looks like  paramagnetic impurity response which quickly
saturates at $\sim2$~T. In the intermediate field region the slope
of the curve is almost linear, indicating some constant background
susceptibility, and close to 14~T a rapid increase of magnetization
is observed, which is a signature of gap closing.

\subsection{ESR}

\begin{figure}
\begin{center}
  \includegraphics[width=0.45\textwidth]{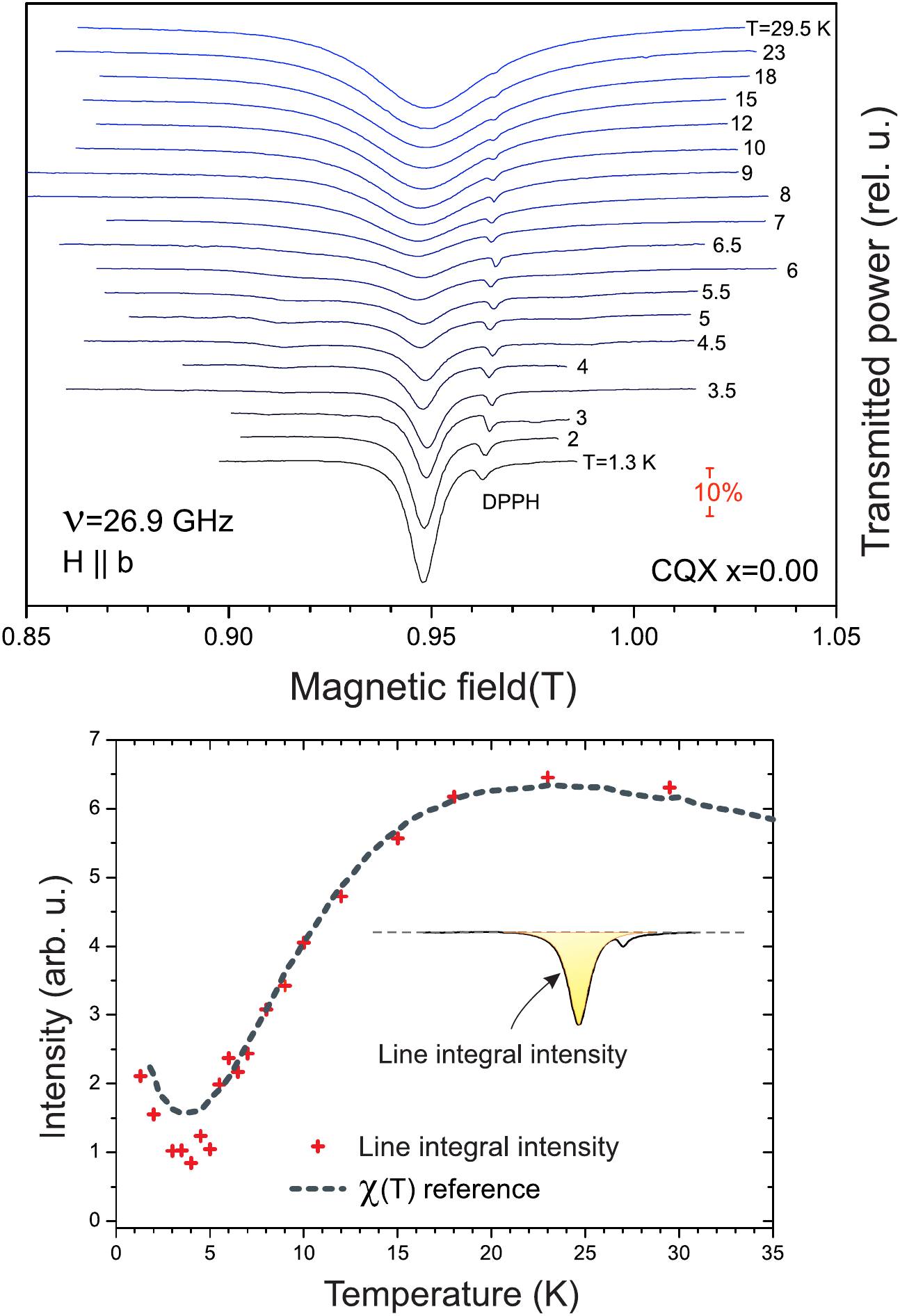}\\
  \caption{Upper panel: normalized power transmission through the cavity with a single crystal of CQC, recorded as a function of magnetic field at different temperatures. The red bar of 10\% absorption defines the scale. Field is applied along $b$, frequency $\nu=26.9$ GHz. Small line on the right labeled as DPPH is a magnetic field standard mark~\cite{dpph} with $g=2.00$. Lower panel: ESR line integral intensity (crosses) as a function of temperature, together with the scaled $\chi(T)$ for a single crystal (dashed line).}\label{FIG:ESR}
\end{center}
\end{figure}

ESR measurements were performed in  a frequency range 25 -- 50 GHz on single crystals of CQC and CQB. Spectra for both at $0\leq H\leq4$~T and $1.3\leq T\leq20$~K consist of a narrow single line ($\Delta H_{1/2}\sim0.01$~T), as shown in Figure~\ref{FIG:ESR} for the CQC case. The principal values of $g$-tensor, obtained at $T=4.2$~K, constitute $g_{b}=2.03$, $g_{ac}^{min}=2.10$, $g_{ac}^{max}=2.28$ for CQC and $g_{b}=2.02$, $g_{ac}^{max}=2.22$  for CQB. This is in good agreement with the average value $\langle g\rangle\simeq2.12$, estimated from disordered powder susceptibility data. The angular dependencies look very conventional, and at such temperature only carry information on the $g-$gactor.

In all cases the observed absorption line gets narrow at low temperatures, while its amplitude increases. As shown in  Figure~\ref{FIG:ESR} (lower panel), the temperature dependence of total intensity follows the temperature dependence of susceptibility in agreement with Kramers--Kronig relations~\cite{WhiteBook}. The high-temperature ESR signal is produced by thermally activated excitations of the spin ladder, while the low-temperature part is due to the impurities. The absorption peak is well described by a single Lorentzian line, with its center independent of temperature within the experimental precision. As the low-field magnetic response of CQX is only due to impurities below $T\sim4$ K, this kind of ESR line shows the impurities to be effectively decoupled $S=1/2$ of copper ions, having the same local surrounding and hence the same $g$-factor as the copper ions, belonging to the ladder structure. This sort of impurities, being found in the same amount in both powder and single crystal samples, is probably originating from structural defects, which lead to effectively broken ladder ends.

\subsection{Quantitative analysis of magnetic data}

In our analysis we have assumed that both $\chi(T)$ and $M(H)$ are quantitatively described by the same ladder model with impurities, and we account for the following factors:

\begin{enumerate}
  \item the ladder system itself, described by parameters $J_{l}$, $J_{r}$, $g$ and $n$, which is the fraction of ladder spins in our system,
  \item impurities, which are considered as free $S=1/2$ with fraction of $n_{imp}=1-n$ and $g_{imp}=g$, as shown by ESR,
  \item magnetic background, which is almost temperature-independent. Sources of this sort of background are both diamagnetic responses of CQX, paraffin and sample holder,
\end{enumerate}

To quantitatively fit the susceptibility, we consider two regimes: the high- (300~K -- 35~K) and low-temperature (35~K -- 2~K) one. For the high-temperature description of a spin ladder system we use high-temperature series expansion (HTSE) by B\"{u}hler \emph{et al.}~\cite{HTSEfromPRB} (please note that our definition of $\alpha$ is inverse to that used in the paper cited). For the low-temperature regime we employed an empirical interpolation function, based on the results of the quantum Monte-Carlo (QMC) simulations, performed with the ALPS package~\cite{ALPS}. As the function definition itself is quite bulky, we refer the reader to \ref{AppendixA} for all the details. Here we only note, that it is based on an empirical interpolation of $\chi(T)$ curves, given by Barnes and Riera~\cite{QMCsuscepInterpolation}. The dimensionless susceptibility per spin $\chi_{QMC}^{*}(t,\alpha)$ is a function of reduced temperature $t=k_{B}T/J_{leg}$ and couplings ratio $\alpha$. It is deduced from QMC and is related to the actual susceptibility per mole of the compound $\chi_{QMC}(T)$ by:

\begin{equation}\label{EQ:ChiQMC}
    \chi_{QMC}(T)=n\frac{(g\mu_{B})^{2}N_{A}}{k_{B}}\chi_{QMC}^{*}\left(\frac{J_{l}t}{k_{B}}, \frac{J_{r}}{J_{l}}\right).
\end{equation}

For the region 35 -- 60~K both high- and low-temperature approaches were used; the resulting curves from HTSE and QMC are almost indistinguishable in this temperature range.

Constructing a fitting function for magnetization is somewhat more involved, as it now includes the magnetic field as a parameter in addition to temperature and $\alpha$. Nonetheless, a proper functional form can be found, as one notices that the actual variable, related to magnetic field is not just $h=g\mu_{B}\mu_{0}H/J_{l}$, but $\xi=\exp(\frac{D-h}{t})$, where parameter $D$ is a function of $\alpha$ only. With a $h\rightarrow\xi$ substitution each single magnetization curve is described as a 2-parameter generalized hyperbola. The hyperbola coefficients can be described as polynomials of $t$ and $\alpha$ (please see \ref{AppendixA} for details). A dimensionless interpolation function $M_{QMC}^{*}(t,h,\alpha)$ is related to the actual fitting function for our magnetization data as:

\begin{equation}\label{EQ:MQMC}
    M_{QMC}(T,H)=ng\mu_{B}N_{A} M_{QMC}^{*}\left(\frac{J_{l}t}{k_{B}},\frac{J_{l}h}{g\mu_{B}\mu_{0}},\frac{J_{r}}{J_{l}}\right).
\end{equation}

The fit was performed by minimizing the weighted square-deviation function $f$ of both $M(H)$ and $\chi(T)$ datasets for each compound:

\begin{equation*}
    \begin{aligned}
        f=\dfrac{1}{N_{\chi}}&\sqrt{\sum\limits_{i=1}^{N_{\chi}}\left(1-\frac{\chi_{f}(T_{i})}{\chi(T_{i})}\right)^{2}}+\\
        +&\dfrac{1}{N_{M}}\sqrt{\sum\limits_{i=1}^{N_{M}}\left(1-\frac{M_{f}(H_{i})}{M(H_{i})}\right)^{2}}.
    \end{aligned}
\end{equation*}

Here $N$ is the number of points in each dataset, $\chi(T_{i})$ and
$M(H_{i})$ are the experimental values and
$\chi_{f}=\chi_{QMC}+\chi_{imp}+\chi_{bkgr}$ and
$M_{f}=M_{QMC}+M_{imp}+\chi_{bkgr}H$ are the final fitting
functions. The resulting curves for $\chi(T)$ and $M(H)$ are shown
together with the data in Figures~\ref{FIG:Susceptibility} and
\ref{FIG:Magnetization}. The insets show the raw data and
$\chi_{f}$, $M_{f}$, while the main panels show the data with the
fitted impurities and background contributions subtracted together
with pure $\chi_{QMC}$, $M_{QMC}$.

\begin{figure}
\begin{center}
  \includegraphics[width=0.5\textwidth]{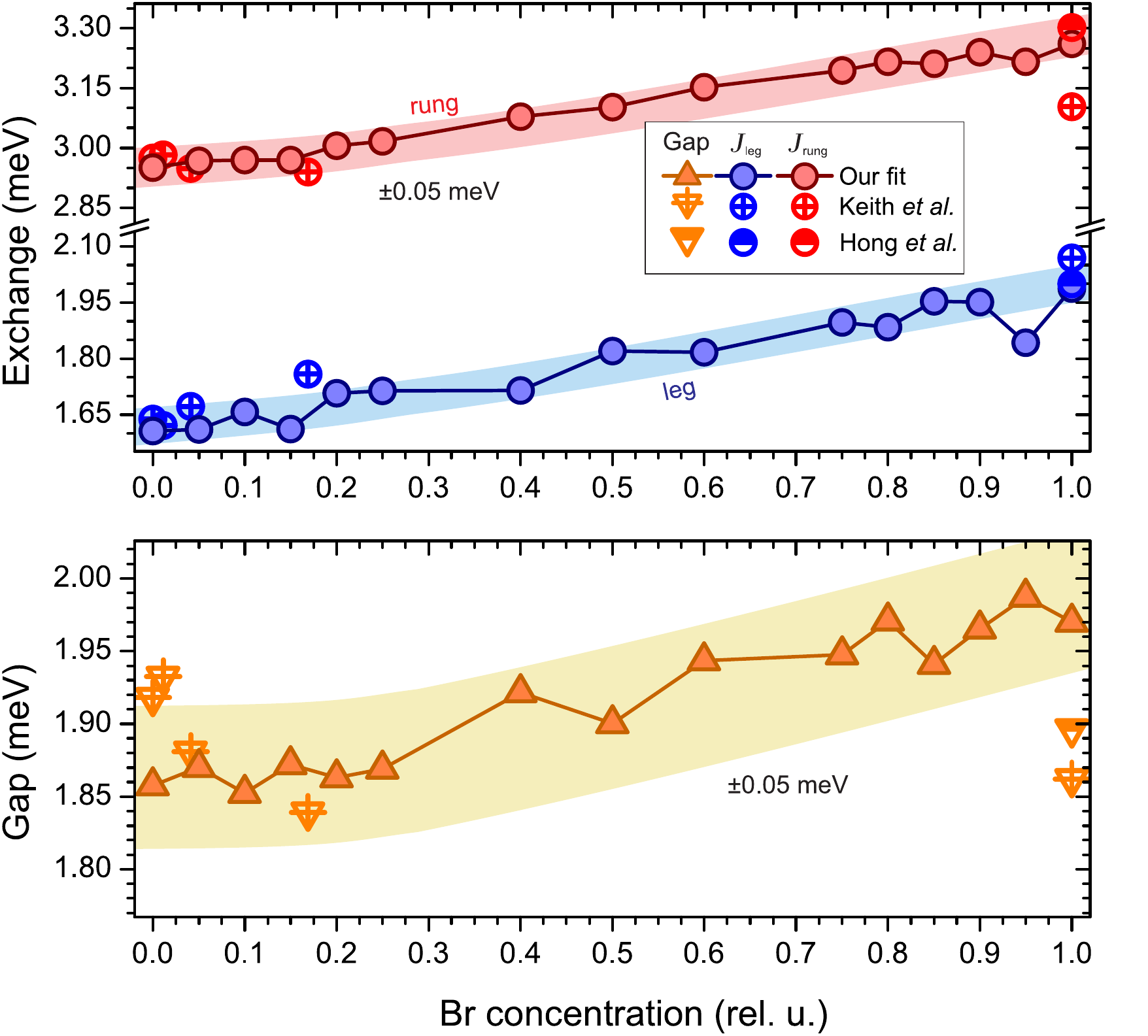}\\
  \caption{Upper panel: $J_{l}(x)$ and $J_{r}(x)$; lower panel: value of $\Delta(x)$. Solid symbols correspond to our result, obtained as described in the text (equations (\ref{EQ:ChiQMC},\ref{EQ:MQMC},\ref{EQ:Gapformula})), hollow-crossed symbols are the result of Keith \emph{et al.} from susceptibility measurement ~\cite{Polyhedron2011}, and half-filled symbols are the result of inelastic neutron scattering experiment~\cite{KenzelPRB}. Shaded bands show the variation in obtained values of $\pm0.5$ K.}\label{FIG:Fitresult}
\end{center}
\end{figure}

The most essential and non-trivial yields of this fit are the strengths of the main exchange interactions $J_{l}$ and $J_{r}$. For $\alpha\geq1$ the gap value is derived from the coupling constants as~\cite{JohnstonArXive}

\begin{equation}\label{EQ:Gapformula}
    \Delta=J_{r}(1-1.0\alpha^{-1}+0.6878\alpha^{-2}-0.1861\alpha^{-3}),
\end{equation}

and the corresponding critical field is $\mu_{0}H_{c}=\Delta/g\mu_{B}$. The results of the fit are presented both numerically and graphically in Table~\ref{TAB:Fitparams} and Figure~\ref{FIG:Fitresult}, respectively. One can see that a noticeable variation of exchange parameters begins as $x$ exceeds 0.25. This is in agreement with the finding by Keith \emph{et al.}~\cite{Polyhedron2011}, that at low substitution Br tends to occupy terminal positions in the Cu$_{2}$X$_{4}$ "group". There is also a correlation with lattice parameter variation present in Figure~\ref{FIG:Cellvariations}: almost no change is observed for small $x$ values.

\begin{table}
\begin{center}
\begin{tabular}{|c||c|c|c|c|c|}
  \hline
$x$ & $J_{l}$ & $J_{r}$ & $n_{imp}$ & $\Delta$ & $\mu_{0}H_{c}$ \\
\hline
rel. u. & meV & meV & \% & meV & T \\
\hline\hline
0.00 & 1.61 & 2.95 & 1.03 & 1.86 & 14.3 \\
0.05 & 1.61 & 2.97 & 1.48 & 1.87 & 14.3 \\
0.10 & 1.66 & 2.97 & 1.12 & 1.85 & 14.2 \\
0.15 & 1.61 & 2.97 & 1.26 & 1.87 & 14.4 \\
0.20 & 1.71 & 3.01 & 1.03 & 1.86 & 14.3 \\
0.25 & 1.71 & 3.02 & 0.77 & 1.87 & 14.4 \\
0.40 & 1.71 & 3.08 & 1.24 & 1.92 & 14.6 \\
0.50 & 1.82 & 3.10 & 0.93 & 1.90 & 14.7 \\
0.60 & 1.82 & 3.15 & 0.90 & 1.94 & 14.9 \\
0.75 & 1.90 & 3.19 & 0.83 & 1.95 & 15.0 \\
0.80 & 1.88 & 3.22 & 0.99 & 1.97 & 15.2 \\
0.85 & 1.95 & 3.21 & 0.71 & 1.94 & 15.0 \\
0.90 & 1.95 & 3.24 & 0.88 & 1.97 & 15.1 \\
0.95 & 1.84 & 3.22 & 0.71 & 1.99 & 15.4 \\
1.00 & 1.99 & 3.26 & 0.91 & 1.97 & 15.3 \\
  \hline
\end{tabular}
\caption{Fit results versus Br concentration $x$.}\label{TAB:Fitparams}
\end{center}
\end{table}

\subsection{Specific heat}

\begin{figure}
\begin{center}
  \includegraphics[width=0.45\textwidth]{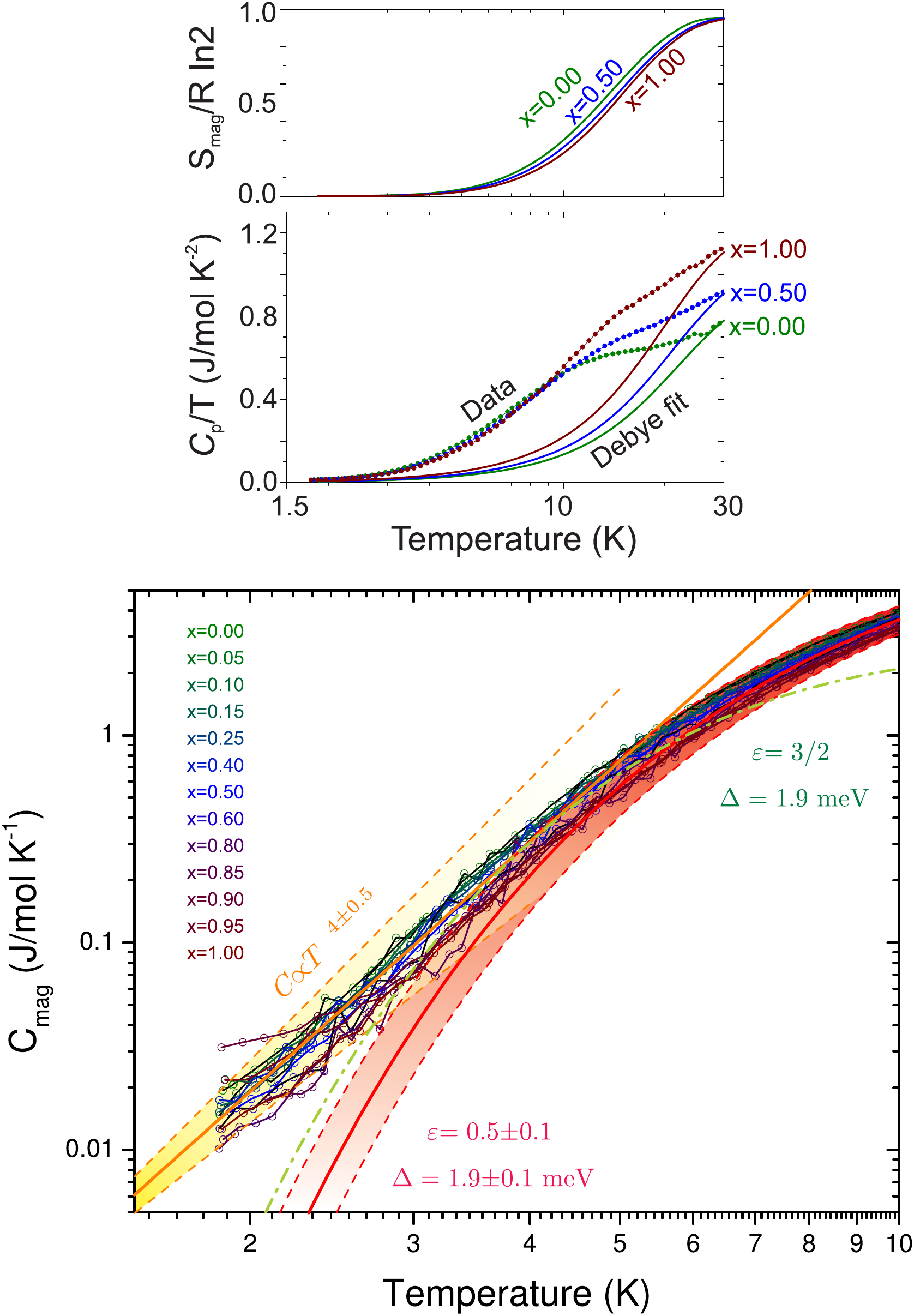}\\
  \caption{Upper panel: full specific heat, estimated lattice contribution and estimated magnetic entropy for $x=0$, 0.5, 1. Lower panel: logarithmic plot of magnetic specific heat for all CQX samples together with the two approximations, given by~(\ref{EQ:HCdecay}) and empirical low-temperature asymptotic (solid and dash-dotted lines correspond to the given value of the parameters, dashed lines correspond to their error boundaries and shaded bands are within these error boundaries).  Color coding for $x$ is the same as at Figures~\ref{FIG:Susceptibility} and~\ref{FIG:Magnetization}.}\label{FIG:Speciheat}
\end{center}
\end{figure}

The specific heat of single CQX crystals with the typical mass of 1-2 mg was measured under zero-field conditions from 1.8 to 200 K. The lattice contribution was subtracted in a Debye approximation with the following assumptions: a) $C_{latt}+C_{mag}\simeq C_{latt}\propto T^{3}$ at the lowest temperatures and b) $\int C_{mag}T^{-1}dT=R\ln2$ --- a constraint, put on the magnetic entropy. The full heat capacity, lattice contributions and corresponding magnetic entropy are present in the upper panel of Figure~\ref{FIG:Speciheat} for $x=0$, $0.5$ and $1$ samples. The low-temperature part of magnetic heat capacity for all the samples measured is shown in the main panel of this figure. It is decreasing rapidly with cooling, but the decrease slows down at the lowest temperatures. While a rapid decrease of $C_{mag}$ is characteristic for a gapped ground state, additional specific heat at low temperatures is to be attributed to the impurity states. The general form of low-temperature specific heat for a gapped system can be written as

 \begin{equation}\label{EQ:HCdecay}
    C_{mag}\propto \left(\dfrac{\Delta}{k_{B}T}\right)^{\varepsilon}\exp\left(-\frac{\Delta}{k_{B}T}\right).
\end{equation}

As it has been shown in~\cite{CpAsymptot}, for the strong-rung limit
of a spin-ladder system $\varepsilon=3/2$. However, our case is
rather intermediate between strong-leg and strong-rung as
$\alpha\sim1$. We find better agreement with the experimental data
for $\varepsilon\simeq0.5$ and the value of the gap
$\Delta\simeq1.9$~meV. This agreement holds in the intermediate
temperature region $k_{B}T\lesssim\Delta$, but breaks down below 4
-- 5~K due to impurities contribution. Between 2 and 4~K this
contribution can be described by a power-law dependence
$C_{m}^{imp}\propto T^{4\pm0.5}$. This power law should be viewed as
no more than an empirical function.

\subsection{Discussion}

As shown above, the magnetic properties of spin-ladder family CQX
can be completely described by a spin-ladder model with paramagnetic
impurities down to $T=2$~K (which is slightly lower that $0.1$ of
the average exchange $\aver{J}/k_{B}$) and up to $\mu_{0}H=14$~T
($g\mu_{B}\mu_{0}H\sim \Delta$). This model works in the whole range
of Br substitution $x$, as well for diluted compounds, as for pure.
Bond disorder in CQX manifests itself only in a gradual
renormalization of exchange constants $J_{l}(x)$ and $J_{r}(x)$. The
amount of impurity spins is always about 1\% and does not show an
obvious dependence on $x$. Moreover, the same amount of impurities
is present in nominally pure compounds, which suggests these
impurities to be related to structural imperfections rather than to
bond disorder. As the $g$-tensor for an impurity spin perfectly
coincides with the $g$-tensor of a ladder spin, we can conclude that
they are in the same local environment. Thus, the most natural
explanation for the origin of impurities is the presence of
structural defects resulting in broken ladder bonds. As it has been
shown in~\cite{RandomizedLadders}, for just 1\% of such an end spins
correlation effects between them are negligible approximately down
to $0.01\langle J\rangle$ in an isotropic ($\alpha=1$) spin-ladder
case and they can be treated as purely paramagnetic. This is also in
agreement with our observations: the magnetic response of the
impurities is well described by the Brillouin function for $S=1/2$.

The effect of halogen substitution on magnetic interactions is well
known for quantum magnets. Such a substitution can strongly affect
both static and dynamic properties of the magnetic
system~\cite{IPA_boseglass, IPA_magnons, CPC_doped, PHCC_critical,
PHCC_phasediagr, CCC_esr1, CCC_esr2, CCB_esr}. In the case of CQX
one would expect the substitution to affect the rung exchange, as it
is created by the bihalide bridging between the Cu$^{2+}$ ions. This
conjecture can also find support in the results of structural
analysis of CQX compounds with different degree of substitution,
which are presented in Figure~\ref{FIG:Cellvariations}. Unit cell
remains almost unchanged in $b$ and $a$ directions for any degree of
substitution $x$, while for $c$ direction a significant change of
$\sim6$\% is observed. A similar change is also found in
Cu--Cl/Br--Cu distances, which belong to the $ac$ plane. This can be
seen as the direct consequence of larger Br ion radius. It is also
interesting to note that almost no change can be found for small
$x$, while for larger substitutions geometry distortion is almost
linear. This is in agreement with the observation of Br ions
tendency to occupy terminal positions in the CQX structure
first~\cite{Polyhedron2011}. Experimental results for the rung
exchange (Figure~\ref{FIG:Fitresult}) show agreement with the
speculations above: $J_{r}$ is obviously affected by the
substitution. Overall increase in $J_{r}$ between CQC and CQB
consists $~10$\%, and this increase also correlates with the bond
geometry change for intermediate values of $x$. The surprising
result is that mediated by the quinoxaline ligand leg exchange
$J_{l}$ is also strongly affected by the substitution, though no
change is observed in the relevant geometry. The change for $J_{l}$
is even more pronounced than for $J_{r}$: it consists about 23\%.
Numerical calculations by Jornet-Somoza \emph{et
al.}~\cite{LandeeInorgChem2012} have shown, that Cl ion has
increased charge localization within the bihalide bridge compared to
Br ion. Thus chlorine to bromine substitution increases orbital
overlap with quinoxaline, increasing the leg exchange.

\section{Conclusions \label{Conclusions}}

We have studied the family of bond-disordered spin-$1/2$ ladders \CQX\ by means of several techniques. Down to moderately low temperatures $T\sim 0.1\aver{J}$ the site-substituted materials are well described in terms of effective parameters $J_{l}(x)$ and $J_{r}(x)$. We have extracted these parameters over the entire range of $0\leq x\leq1$ from susceptibility and magnetization curves and found them to vary continuously. Thus \CQX\ can be described as a 'tunable' spin ladder material.

\section*{Acknowledgements}

The work at ETH~Z\"{u}rich was in part supported by the Swiss National Fund, Division 2. The work at P.~L.~Kapitza Institute was supported by RFBR grant 12-02-31220. The authors thank V.~N.~Glazkov for stimulating comments and discussions.

\appendix
\section{Fitting functions \label{AppendixA}}

The dimensionless magnetic susceptibility per spin of the spin ladder can be described by an empirical interpolation function~\cite{QMCsuscepInterpolation}

\begin{equation}\label{EQ:ChiQMCr}
\begin{aligned}
    \chi&_{QMC}^{*}(t,\alpha)=\dfrac{1}{4t}\left(1+\left(\frac{t}{\tau_{1}}\right)^{\gamma_{1}}\left(e^{d/t}-1\right)\right)^{-1}\cdot\\
    &\cdot\left(1+\left(\frac{\tau_{2}}{t}\right)^{\gamma_{2}}\vphantom{\left(\frac{t}{\tau_{1}}\right)^{\gamma_{1}}}\right)^{-1}.
\end{aligned}
\end{equation}

The parameters of this function depend on the couplings ratio $\alpha$ as

\begin{equation}\label{EQ:Chicoeffexpansion}
\begin{aligned}
    &\tau_{1}=9.4025-10.9424\alpha+3.9308\alpha^{2};\\
    &\gamma_{1}=1.5421-1.6279\alpha+0.45282\alpha^{2};\\
    &d=-0.032304+0.39862\alpha+0.055093\alpha^{2};\\
    &\tau_{2}=1.7308-1.3808\alpha+0.40998\alpha^{2};\\
    &\gamma_{2}=2.3945-1.3799\alpha+0.59455\alpha^{2}.
\end{aligned}
\end{equation}

The interpolation~(\ref{EQ:Chicoeffexpansion}) is valid for $1\leq\alpha\leq2$.

As it was mentioned in the main text, the magnetization per spin is well described by a generalized hyperbola $F(\xi)=B_{1}+B_{2}/(1-\xi)^{\gamma}$, where $\xi=\exp(\frac{D-h}{t})$. Considering that $M=0$ at $H=0$ we end up with the interpolation function form

\begin{equation}\label{EQ:MQMCr}
    \begin{aligned}
        M&_{QMC}^{*}(t,h,\alpha)=\\
        &=B\left[(1-e^{\frac{-D}{t}})^{-\gamma}-(1-e^{\frac{-D+h}{t}})^{-\gamma}\right].
    \end{aligned}
\end{equation}

The dependencies of the parameters on $t$ and $\alpha$ were found by fitting the QMC data in the range $1\leq\alpha\leq2$, $0\leq h\leq1.2$ and $0.07\leq t\leq0.12$. We obtained:

\begin{equation}\label{EQ:Mcoeffexpansion}
\begin{aligned}
    &B=-0.12211+1.946t+0.27165\alpha-0.72784\alpha t\\
        &\phantom{ABCD}-2.0823t^{2}+1.8664\alpha t^{2}-0.093364\alpha^{2};\\
    &D=-0.36317+0.81397\alpha;\\
    &\gamma=0.14404+0.62707t-0.20613\alpha+0.54018\alpha t\\
        &\phantom{ABCD}+0.085184t^{2}-4.6604\alpha t^{2}+0.1315\alpha^{2}.
\end{aligned}
\end{equation}

The response of impurities is described by the $S=1/2$ Brillouin function, for both the magnetization

\begin{equation}\label{EQ:Mimp}
    \begin{aligned}
        M&_{imp}(T,H)=n_{imp}g\mu_{B}N_{A}\cdot\\
        &\cdot\left[2\coth\left(\frac{g\mu_{B}\mu_{0}H}{k_{B}T}\right)-\coth\left(\frac{g\mu_{B}\mu_{0}H}{2k_{B}T}\right)\right]
    \end{aligned}
\end{equation}

and the susceptibility

\begin{equation}\label{EQ:Chiimp}
    \chi_{imp}(T)=M_{imp}(T, H)/H,
\end{equation}

with $\mu_{0}H=0.1$~T.

\bibliography{CQXpaper_Povarov_v4.0}
\bibliographystyle{apsrev4-1}

\end{document}